# Detecting Network Security Vulnerabilities and Proactive Strategies to Mitigate Potential Threats

Aiman Al-Sabaawi*[1], Thamer A. Alrowidhan[2]
[1]School of Computer Science, Queensland University of Technology, Brisbane, Australia
[2]Information and Communications Technology, National Unified Procurement Company, Riyadh, Saudi Arabia
a.alsabaawi@hdr.qut.edu.au, tarowidhan@nupco.com

## Abstract

In multi-tier network systems, custom applications, Web services and platform environments, storing data and information assets becomes a challenge for any organisation. Although there are different methods to secure network systems, the best way to test the level of security is to conduct penetration testing. In this paper, we describe how we performed live penetration testing for a particular network, namely, 192.168.3.0/24 (Case Study) by identifying the system vulnerabilities to enable its penetration. After compromising the system, critical data (Flags) must be found, indicating our successful penetration. As professional penetration testers, we used an arsenal of penetration testing tools utilised by malicious actors on the internet, such as Nmap, Nessus, Sparta and Metasploit, etc. Typically, much effort was employed on reconnaissance & scanning phases, rather than system exploration, due to their importance in identifying security vulnerabilities in the system environment. The vulnerability analysis highlighted the most critical threats, which token is an advantage to gain access, namely, FTP services, HTTP, and human errors. However, comprising the system is not sufficient because the critical data "Flag" generally requires the administrator's rights. Consequently, teams often examine the system to find a way to escalate privilege to the root level. Furthermore, some critical data (Flags) require decryption algorithms or the analysis of captured packets to make them readable. We found eight Flags and identified a system security breach. Mitigation strategies addressing the identified vulnerabilities are recommended to ensure the given networks are secured against future attacks.

## I. INTRODUCTION

The team followed the most professional methods in conducting Penetration Testing. Five phases should be considered, beginning with a reconnaissance phase, to collect passive information about the targeted system. Next, scanning phases were used to collect active information such as open services and open sources. This involved the employment of recognition and scanning tools such as Nmap, Nessus, etc. These are the main phases in conducting successful Penetration Testing. After completing those two phases, we exploited the targeted system using different exploitation tools. Once the target has been compromised, such access must be maintained using different techniques such as creating a back door. Although the clearing tracks phase of professional Penetration Testing was not used in this Penetration Testing assignment, it is essential that such a phase be conducted by white hackers to evaluate the customer logging systems [1][2]. Figure 1 shows the five phases of Penetration Testing.

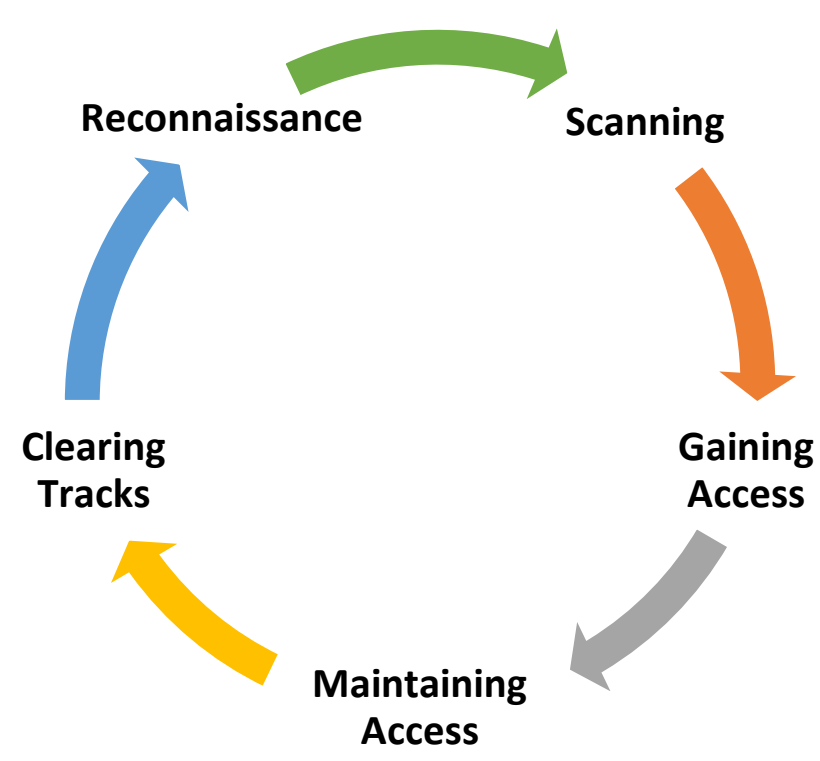

Figure 1. Phases of penetration testing.

# II. RECONNAISSANCE AND SCANNING PHASES

Reconnaissance is the act of collecting primary data or intelligence of targeted victims. The gathered data can guide us through the overview of the network and exploration variabilities on targeted clients. In this regard, the penetration testing team used the most potent scanning tools to detect significant information for reconnaissance of the network, following the steps below [2].

Firstly, Nmap was used for port scanning. Powerful tools such as Nessus and Sparta then performed deeper scanning, including port scanning and OS detections. They also provided a summary of the running services, service vulnerabilities and valuable information to attack the targeted host. As a result, three hosts were found to be live, 192.168.3.(222/111/77). With further deep scanning, another host was found (192.168.3.74) by scanning the network without pinging. This can bypass the host firewall if it exists. The result of target system scanning is shown in Figure 2. Group11 represented our team in these challenges in this paper.

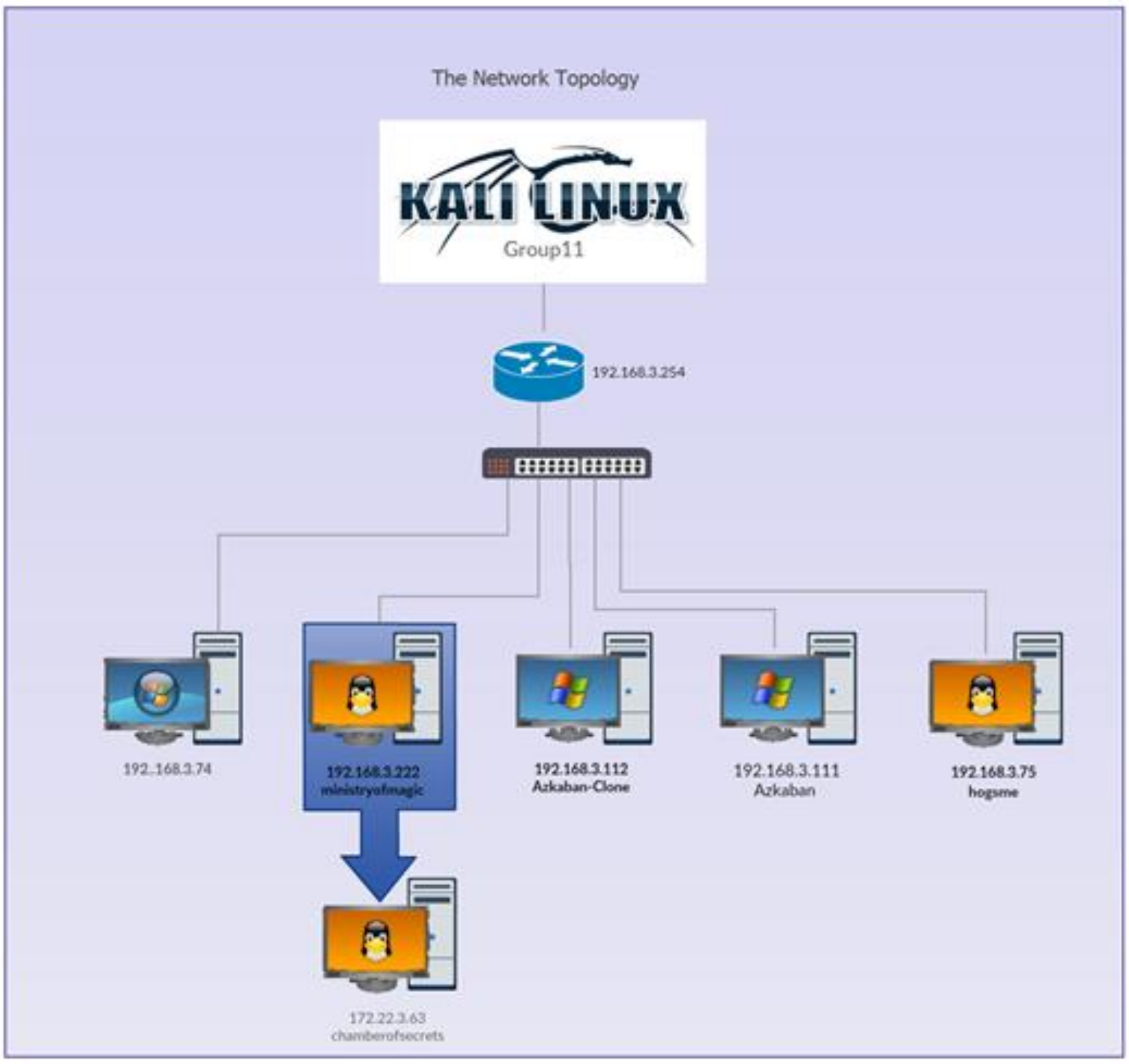

Figure 2. The network topology.

After those processes, the vulnerability analysis tools were used to collect more information regarding live hosts. Sparta and Nessus tools are valuable tools for collecting information and analysing system vulnerabilities. The essential findings obtained from the vulnerability scanner phase are shown in Table 1:

| IP/Host Name | Port | Service | Service Version | Deep Scan Info. |
|---|---|---|---|---|
| 192.168.3.75 hogsme | 21 | FTP | ProFTPD 1.3.3c | -login: guest password: guest |
| | 22 | SSH | OpenSSH 5.9p1 Debian5ubantu1.8 | |
| 192.168.3.111 azkaban 192.168.3.112 azkaban-clone | 135 | TCP | Microsoft EPMAP (End Point Mapper) | MS Windows Server 2003 SP1 or SP2 |
| | 445 | TCP | NetBIOS | |
| | 1027 | TCP | MS Windows RPC (IIS) | |
| | 2016 | TCP | bootserver | |
| | 3389 | TCP | Ms terminal service | |

| IP/Host Name | Port | Service | Service Version | Deep Scan Info. |
|---|---|---|---|---|
| 192.168.3.222 ministryofmagic | 80 | HTTP | Apache httpd2.2.20 Ubantu11 Webcalendar 1.2.4 | Apache -/install/ /log.html /tests/ /tools/ /login/ |
| | 22 | SSH | openSSH 5.8p1Debain7ubantu1 | |
| 192.168.3.74 | 21 | FTP | FTP serive konica Minolta FTP utility 1.0 download | -login: anonymous password: anonymous |

Table 1. Vulnerabilities and Hosts

# III. GAINING AND MAINTAINING ACCESS PHASE

## A. Host 192.168.3.75

### 1) Attack attempts FTP service:

Many attacking attempts were conducted to compromise such hosts using the open services found in the reconnaissance phase. The FTP service is opened using proFTPD (version 1.3.3C) and the operating system was found to be Linux Debian, as shown in Figure 2.

### 2) FTP (proFTPD 1.3.3C) vulnerabilities:

This version is prone to be abused by hackers due to a backdoor in some of these versions, allowing the attacker to remotely access the system with root privileges. Therefore, we started with this version to exploit such a vulnerability, hoping to compromise the system.

### 3) Exploit process FTP service:

The team began by compromising the host using the Armitage exploit tool. This tool has some exploits which take advantage of such vulnerable services. Nevertheless, mostly it uses a back door in the running service. Thus, proftpd_133c_backdoor was used to launch the attack [3]. Access to the host was successful and the interact shell session was created, as shown in Figure 3.

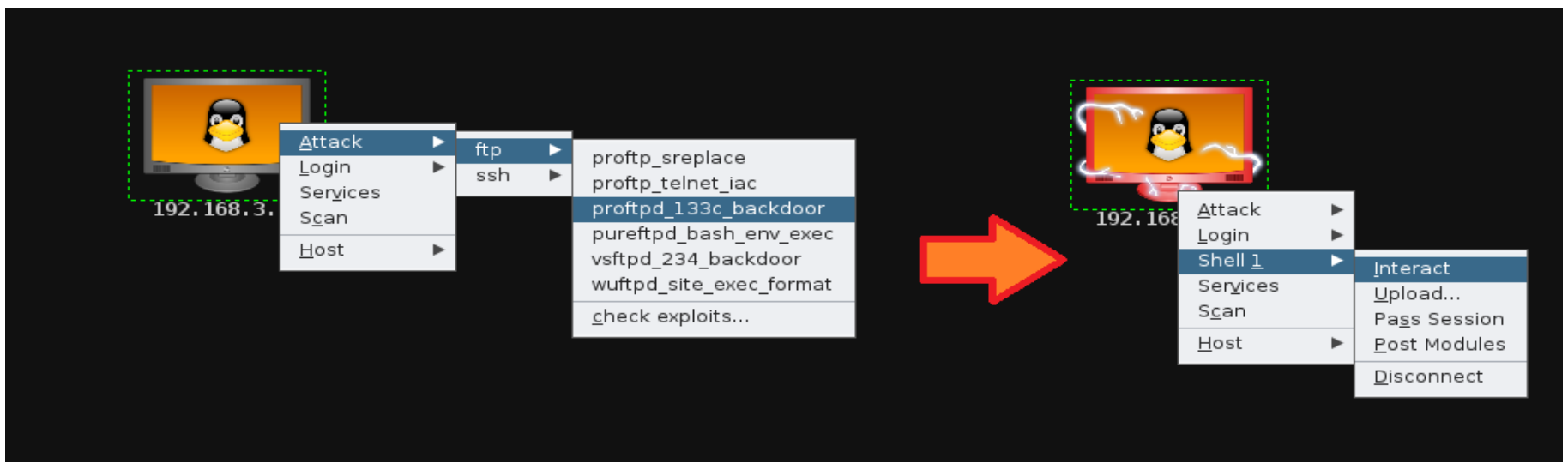

Figure 3. Backdoor exploit on host 192.168.3.75.

Once the shell had been created, the (id) command was issued to identify the privilege that allowed access. A remote root privilege was found. A root privilege permits a user to be added to maintain access to SSH and reset the root password. After that, the standard SSH was opened and the root privilege was shifted. The locate command was then issued to identify the flag files.

### 4) FLAGs Findings:

Two flags were found; one was a hidden file. The first flag was easy to discover by issuing a Cat command to read the file, as shown in Figure 4. The second flag was a hidden file located in the path directory /home/hagrid/flag.txt. A Cat command was issued to read the file and the second FLAG was found, as shown in Figure 5.

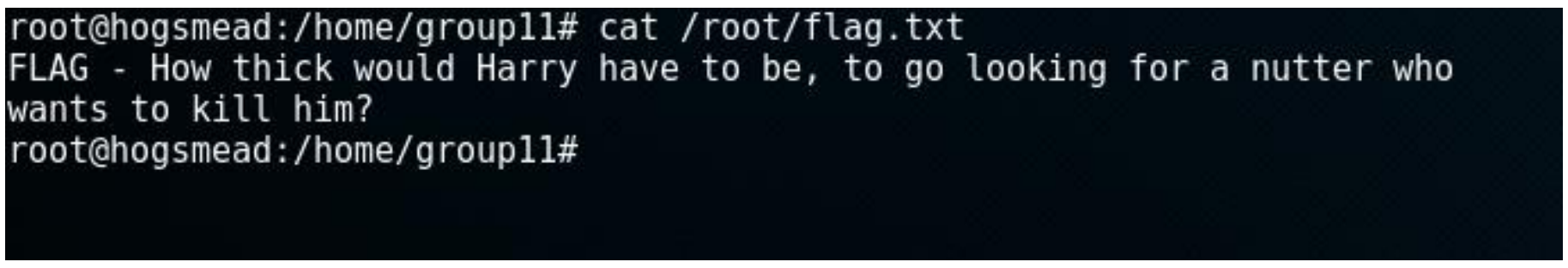

Figure 4. Flag 01.

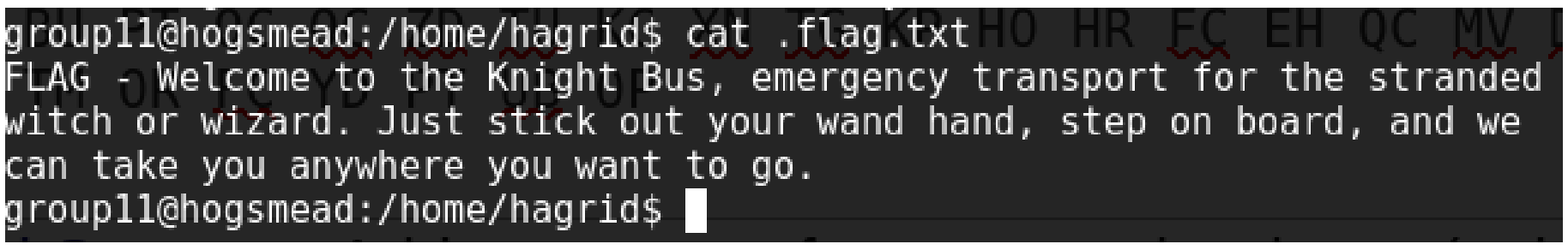

Figure 5. Flag 02.

Another two files were found; the first was named "decrypt me", which has a cipher message. However, another file was a hint file containing a hint text message "the good guys always play fair", as shown in Figure 6. Such a hint was helpful to decrypt the previous cipher message. After searching in google, we found an old cipher named "PlayFair", used in World War Two (WWII). A "Playfair" cipher used a shared key to decrypt this file, translating the letter 'J' Into 'I'. The shared website was used for online decryption. As a result, the third flag was found, as shown in Figure 7.

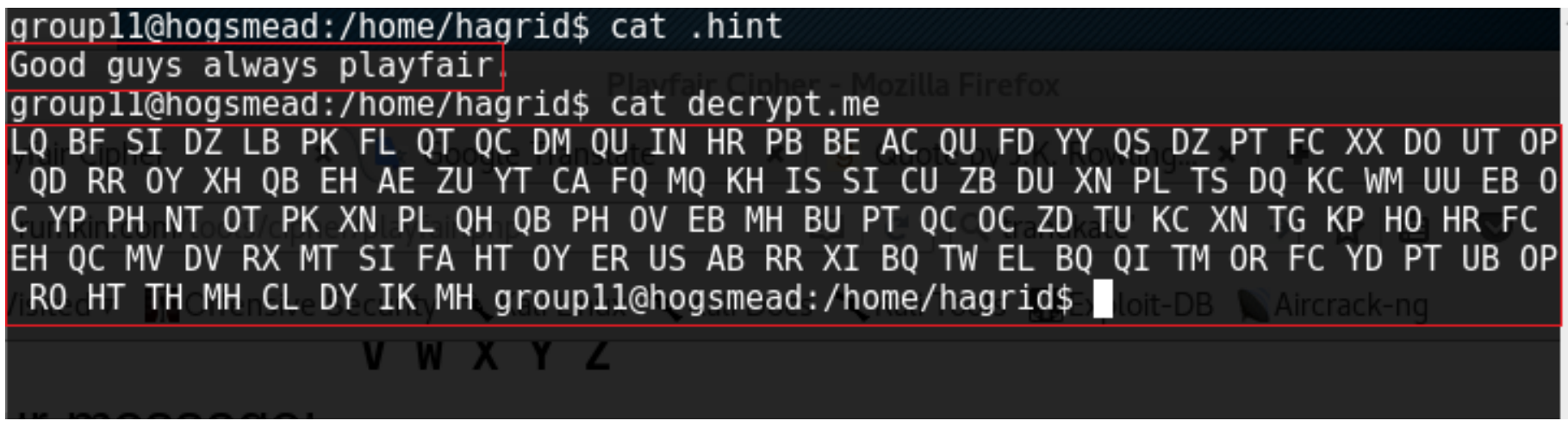

Figure 6. Hint file and decrypt the message.

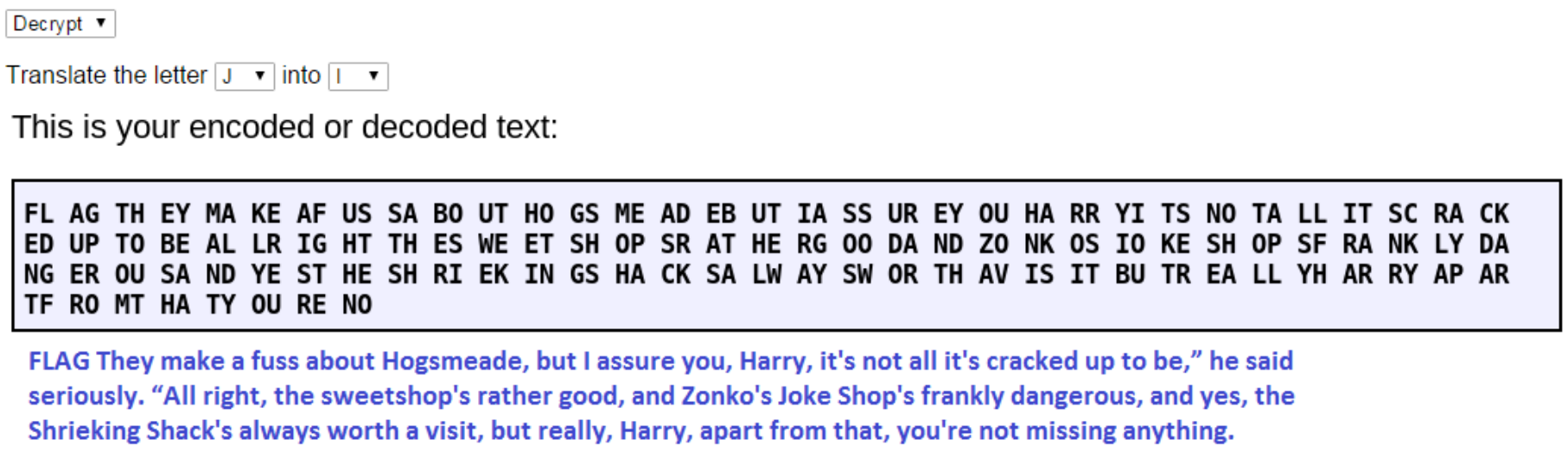

Figure 7. Flag 03.

## B. Host 192.168.3.111

### 1) Attack Attempts Windows server MS08-067 exploit:

From scanning, the host was found to be running on Windows server 2003 SP1 OS, and its remote desktop was opened. This service can exploit vulnerabilities in Windows samba service (remote desktop service), using the Armitage and Metasploit tool [4].

### 2) Exploit Process:

To gain access to this host, Carlos Perez's 'getgui' script was utilised in the Meterpreter shell, which provides access to the remote desktop. The user account and password were then created to gain login access. Thus, after creating a login using the following command, "run getgui –u group11 –p group11", the system using ms08_067_netapi was illustrated, as shown in Figure 8.

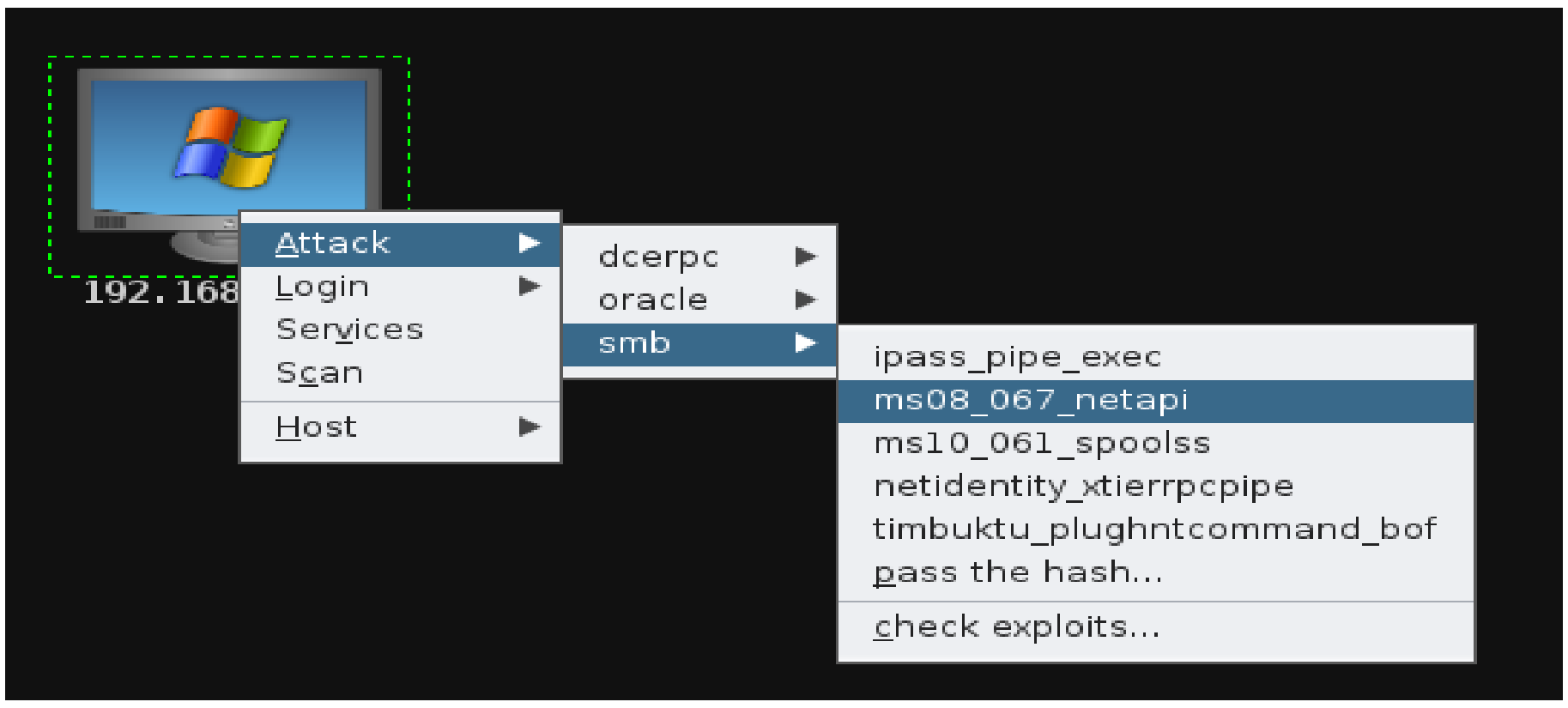

Figure 8. Attack attempts to host 192.168.3.111.

Next, the rdesktop command was run with a created password to gain access to the host and fully compromise it, as shown in Figure 9.

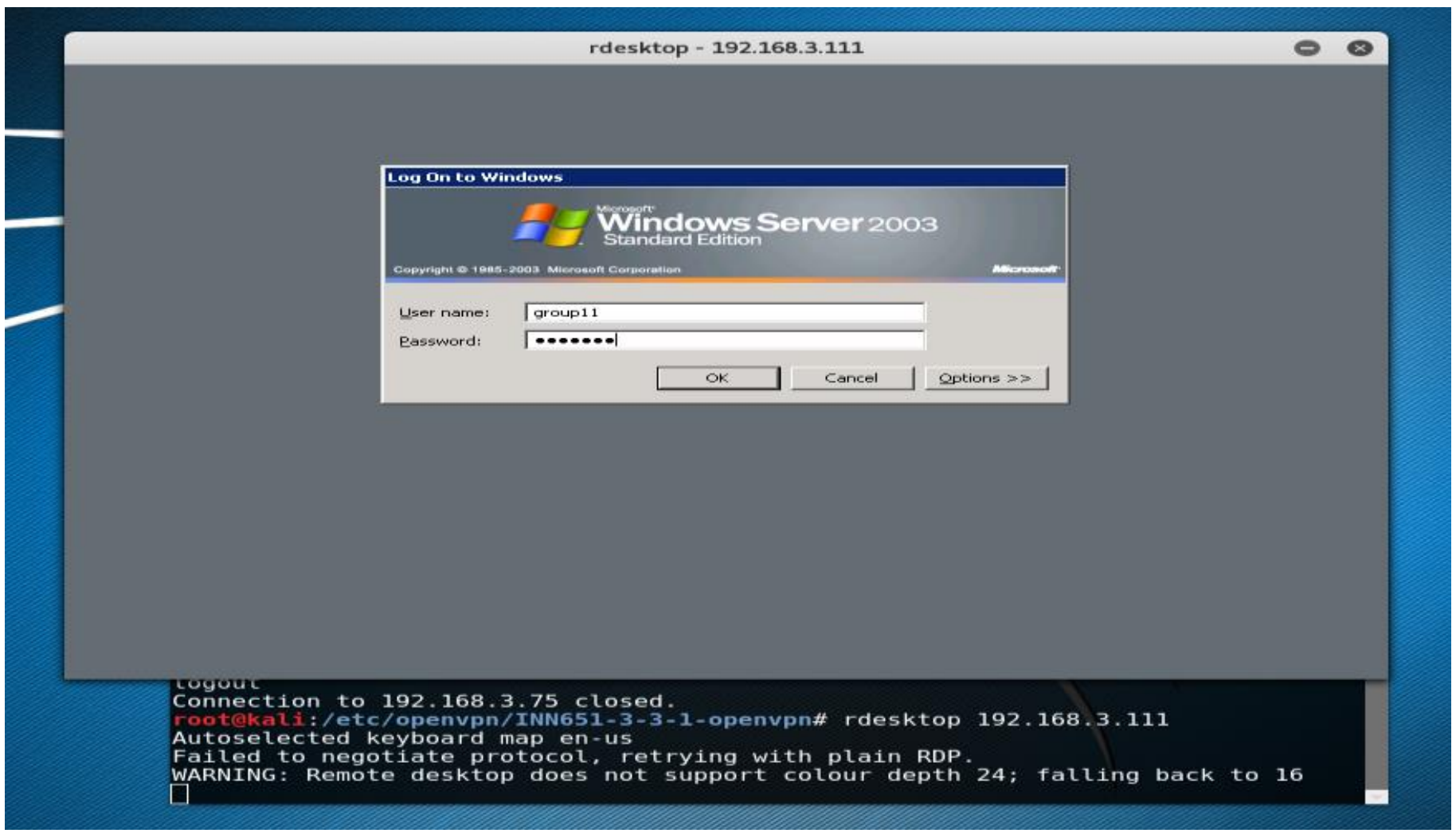

Figure 9. Accessing the host 192.168.3.111 by using the rdscktop command.

### 3) FLAGs finding:

The fourth flag was a text file in the administrator's desktop folder, found by searching Windows files, as shown in Figure 10.

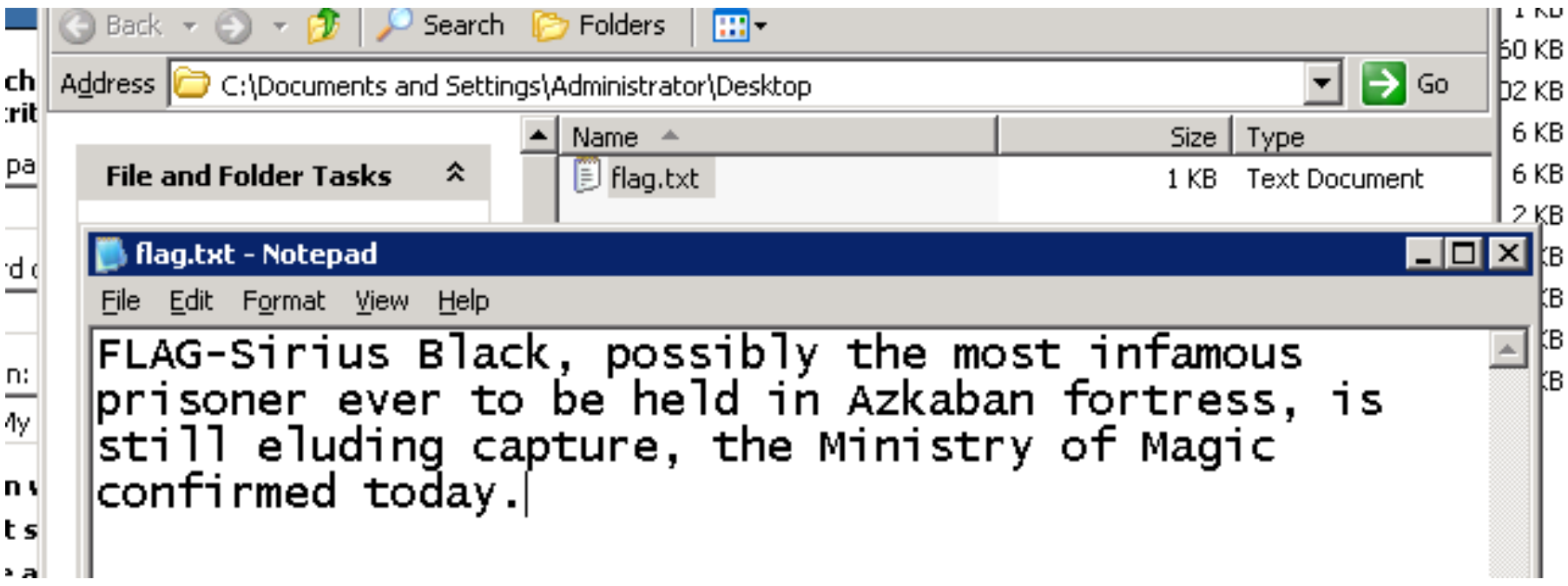

Figure 10. Flag 04.

The fifth flag in the Windows machine is an executable file located in c:\Documents and setting\black\My Documents file. The file needs an administrator's privilege to execute it, so the admin password was reset. Running the "flag.exe" file revealed a small puzzle, so OllyDbg software was used to decode it as assembly language. As a result, the flag file was decoded and the flag was found as static message in code at the end of the puzzle, as shown in Figure 11.

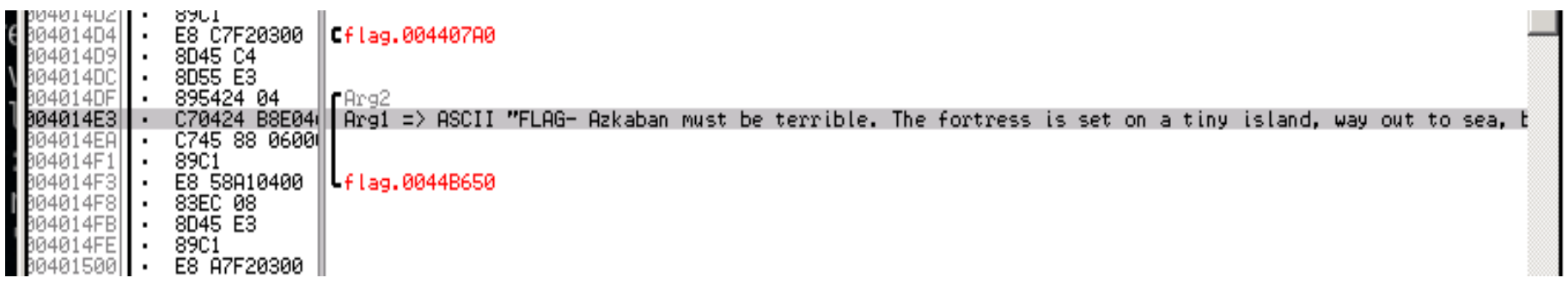

Figure 11. Flag 05.

## C. Host 192.168.3.112

### 1) Attack Attempts (Remote Desktop):

In host 192.168.3.111, there was a hint file with the username "black" and password "regulus" for the clone machine 192.168.3.112. Such findings allowed us to access the clone 192.168.3.112, as shown in Figure 12.

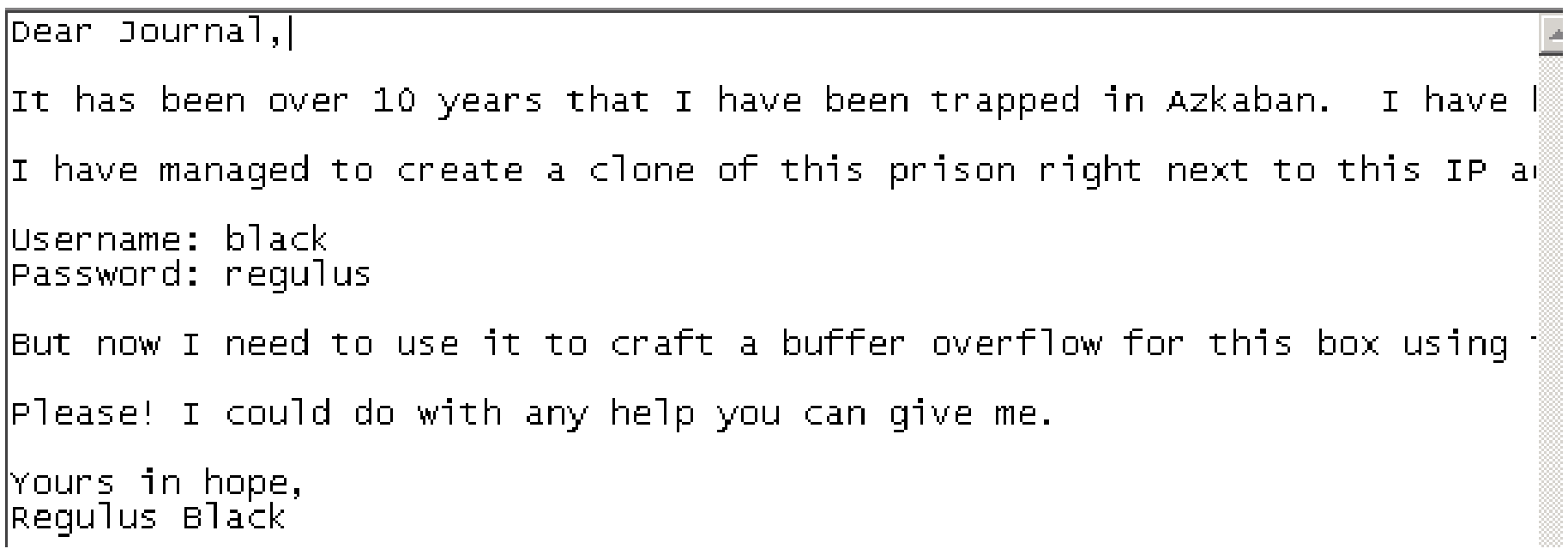

```
Dear Journal,

It has been over 10 years that I have been trapped in Azkaban.  I have l

I have managed to create a clone of this prison right next to this IP a

Username: black
Password: regulus

But now I need to use it to craft a buffer overflow for this box using

Please! I could do with any help you can give me.

Yours in hope,
Regulus Black
```

Figure 12. The message is found in the document user black folder.

### 2) FLAGs findings:

Unfortunately, no Flags were found because the machine was the clone for host 192.168.3.111. However, this host has a variety of programs that are used on host 192.168.3.111.

## D. Host 192.168.3.74

### 1) Attack attempts (FTP service):

Many attacking attempts were conducted to compromise such hosts using the open services found in the reconnaissance phase. For example, when the FTP service was opened, the Konica Minolta FTP Utility was found and the operating system was found to be Windows 2008 server, as shown in Figure 13.

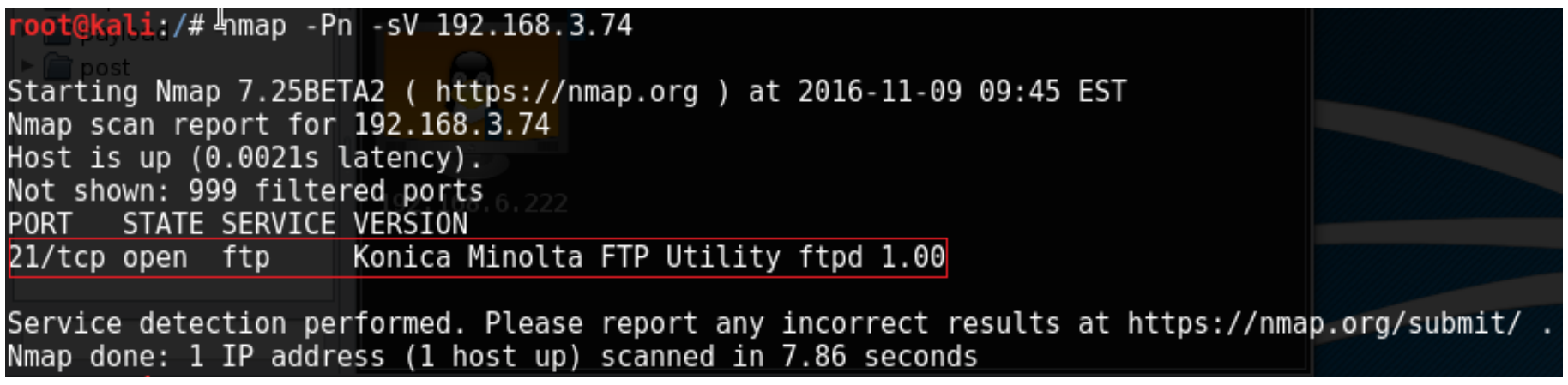

```
root@kali:/# nmap -Pn -sV 192.168.3.74

Starting Nmap 7.25BETA2 ( https://nmap.org ) at 2016-11-09 09:45 EST
Nmap scan report for 192.168.3.74
Host is up (0.0021s latency).
Not shown: 999 filtered ports
PORT   STATE SERVICE VERSION
21/tcp open  ftp     Konica Minolta FTP Utility ftpd 1.00

Service detection performed. Please report any incorrect results at https://nmap.org/submit/ .
Nmap done: 1 IP address (1 host up) scanned in 7.86 seconds
```

Figure 13. The decoded message for Flag 06.

### 2) FTP (Konica Minolta FTP service) vulnerabilities:

A Konica Minolta FTP vulnerability allows remote attackers to run arbitrary commands, so such vulnerabilities were utilised in an attempt to gain access. The file named (39719.ps1) was uploaded, as shown in Figure 14. The purpose of the code is to allow Privilege Escalation and this code can also gain administrator privilege in Microsoft Windows 7, 10 & Server 2008, 2012 (x86/x64). Nevertheless, the exploit which was used did not work, so we moved to the next host.

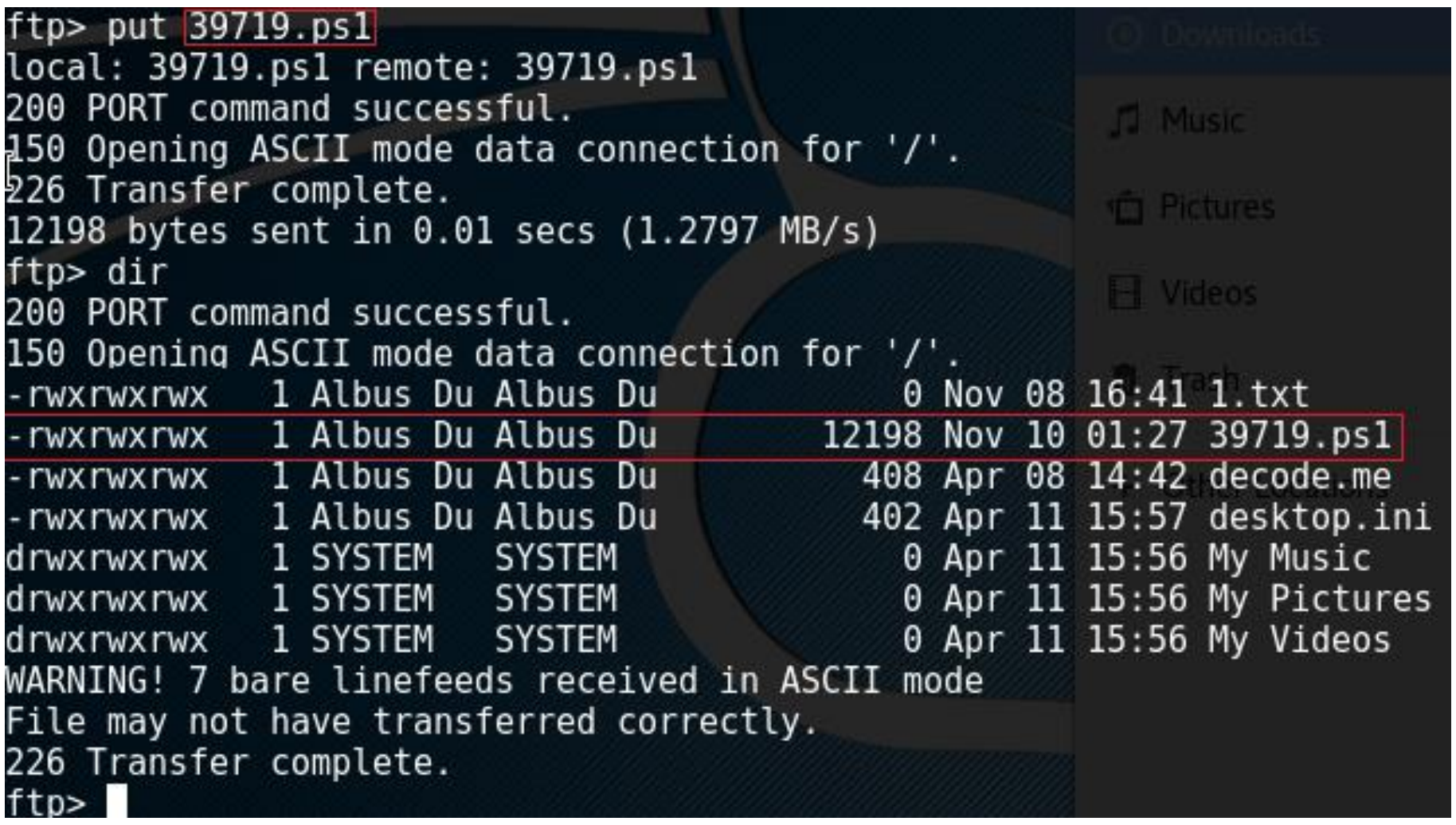

```
ftp> put 39719.ps1
local: 39719.ps1 remote: 39719.ps1
200 PORT command successful.
150 Opening ASCII mode data connection for '/'.
226 Transfer complete.
12198 bytes sent in 0.01 secs (1.2797 MB/s)
ftp> dir
200 PORT command successful.
150 Opening ASCII mode data connection for '/'.
-rwxrwxrwx   1 Albus Du Albus Du            0 Nov 08 16:41 1.txt
-rwxrwxrwx   1 Albus Du Albus Du        12198 Nov 10 01:27 39719.ps1
-rwxrwxrwx   1 Albus Du Albus Du          408 Apr 08 14:42 decode.me
-rwxrwxrwx   1 Albus Du Albus Du          402 Apr 11 15:57 desktop.ini
drwxrwxrwx   1 SYSTEM   SYSTEM              0 Apr 11 15:56 My Music
drwxrwxrwx   1 SYSTEM   SYSTEM              0 Apr 11 15:56 My Pictures
drwxrwxrwx   1 SYSTEM   SYSTEM              0 Apr 11 15:56 My Videos
WARNING! 7 bare linefeeds received in ASCII mode
File may not have transferred correctly.
226 Transfer complete.
ftp>
```

Figure 14. Upload file 39719.ps1.

## 3) Exploit Process Konica Minolta FTP service:

Because the host was compromised using regular FTP access, access was gained to the host using the default FTP password "anonymous" as a guest user. Access to the system was successful, with the FTP's home folder having full privileges. Such privileges allow any user to execute any uploaded file. The file "decode.me" was found with a ciphertext message that may contain the flag. The findings are illustrated in Figure 15.

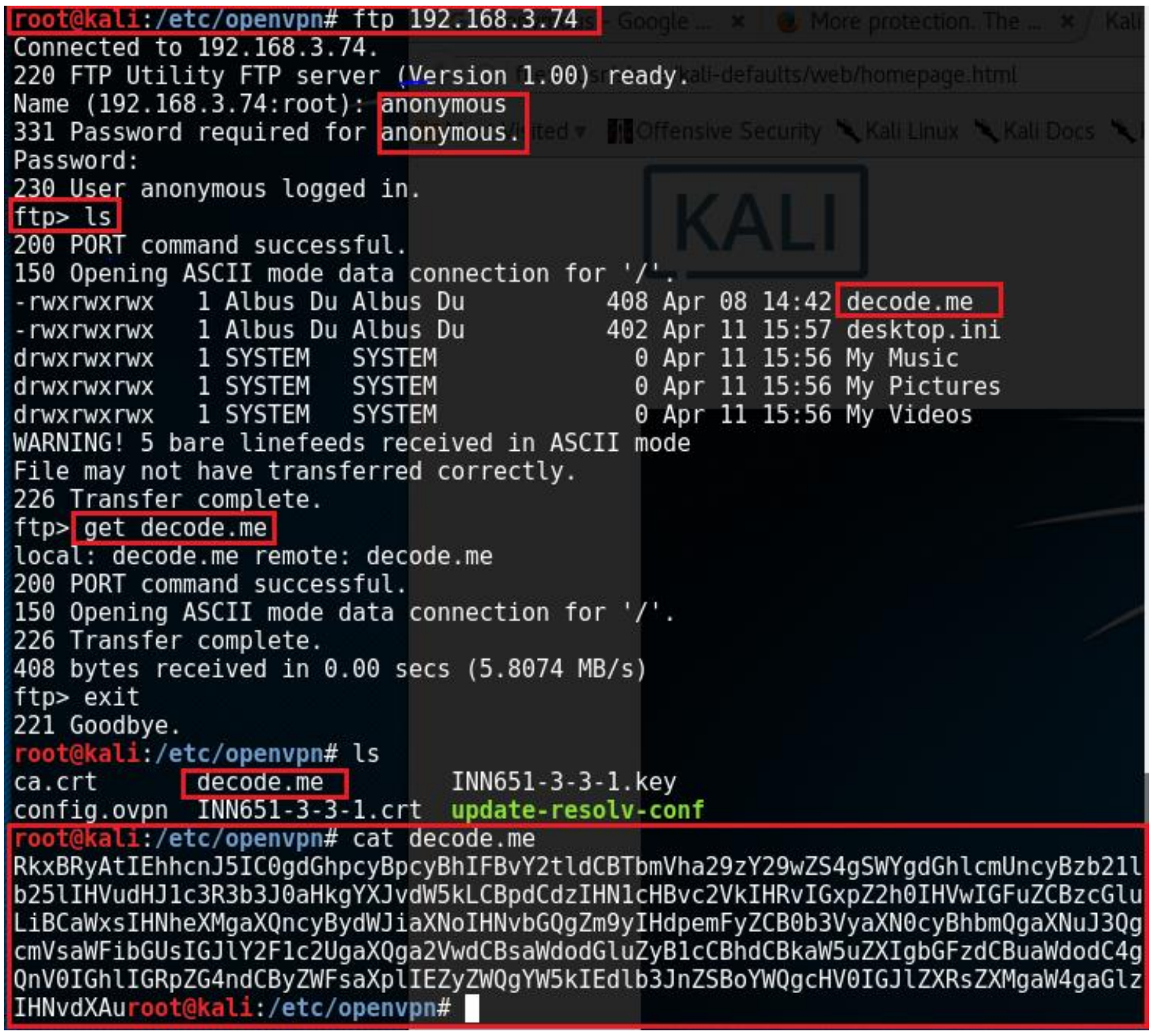

Figure 15. The decoded message for Flag 06

## 4) FLAG Findings:

The sixth flag's message was found after decoding the "decode.me" text message using base 64 decoders, as shown in Figure 16.

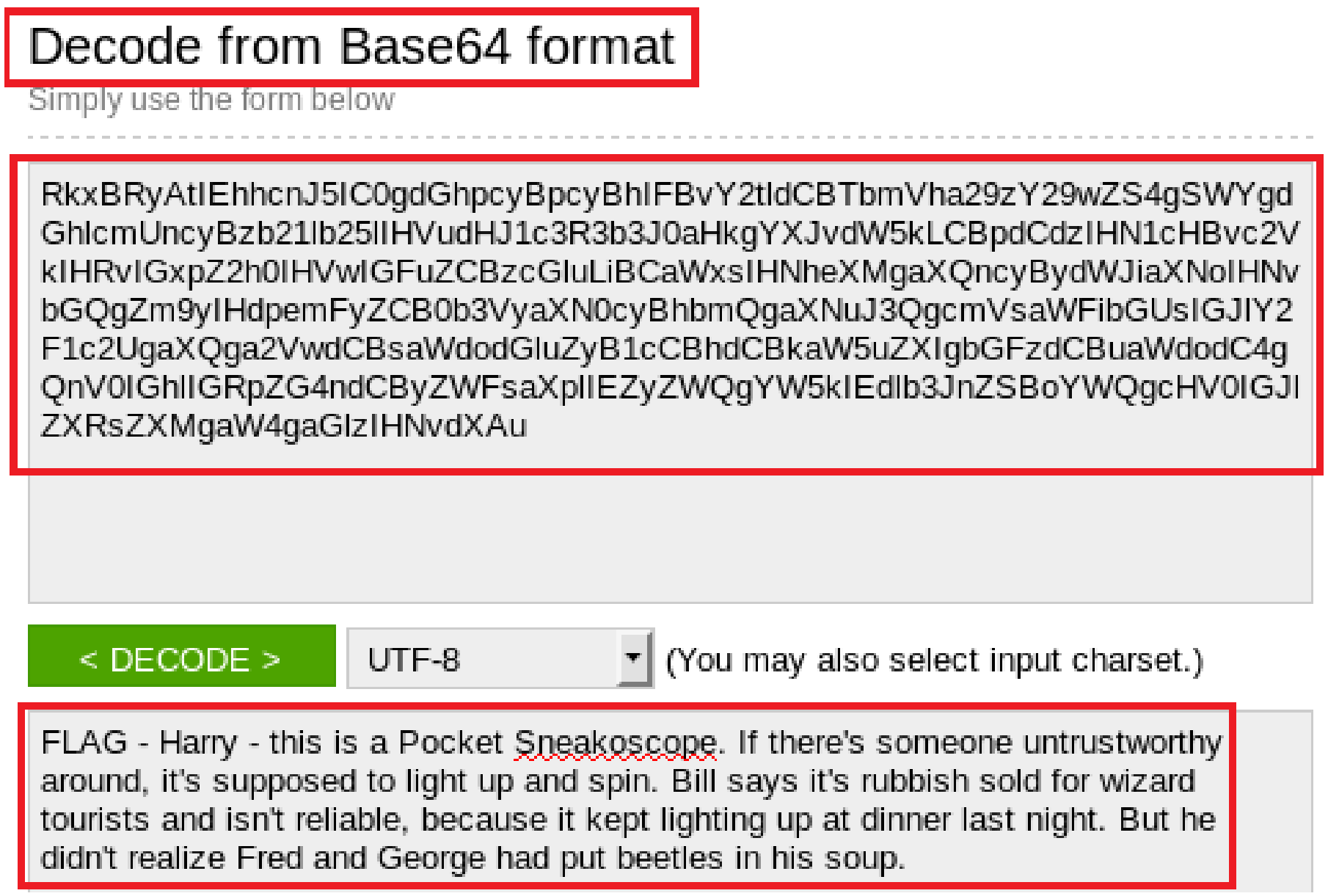

Figure 16. Flag 06 after decoding.

## E. Host 192.168.3.222

### 1) Attack attempt HTTP service:

From the scanning phase, such a host has HTTP services found to be running Webcalendar 1.2.4 web applications which are vulnerable to some exploits. Thus, this vulnerability was taken advantage of to gain access to the host.

### 2) FTP (Webcalendar 1.2.4) vulnerabilities:

Webcalendar 1.2.4 is a vulnerable version regarding an input validation error in sending the "reminder.php" script, which hackers have exploited.

### 3) Exploit process HTTP service:

A Webcalendar exploit was used against host victim 192.168.3.222 to land a Linux shell written in Perl. Therefore, the first step was finding the Webcalendar exploit, as shown in Figure 17.

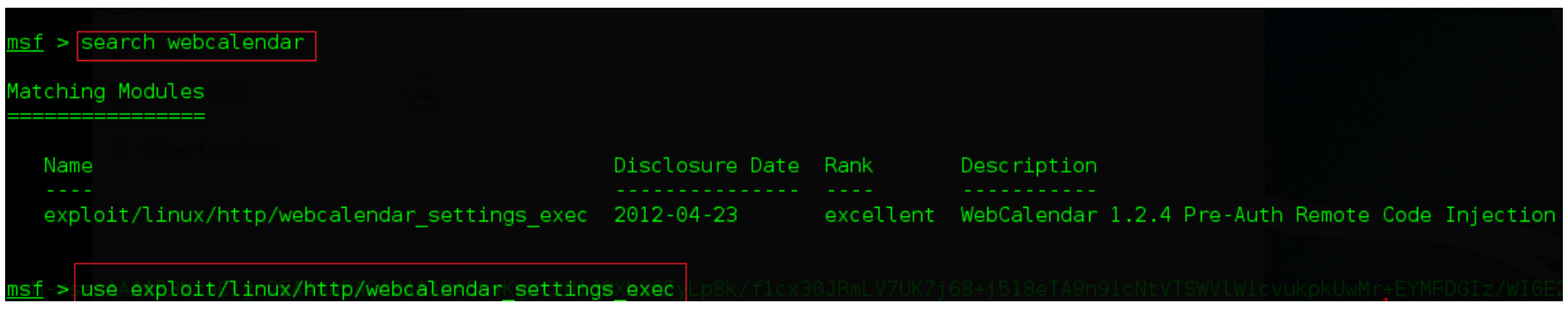

Figure 17. Flag 07 search Webcalendar.

The following procedure ran the selected exploit after setting the suitable option, as shown in Figure 18. The attack was then launched by entering the exploit -J command. After running the attack successfully, the session state was seen by using the show sessions command, as shown in Figure 18.

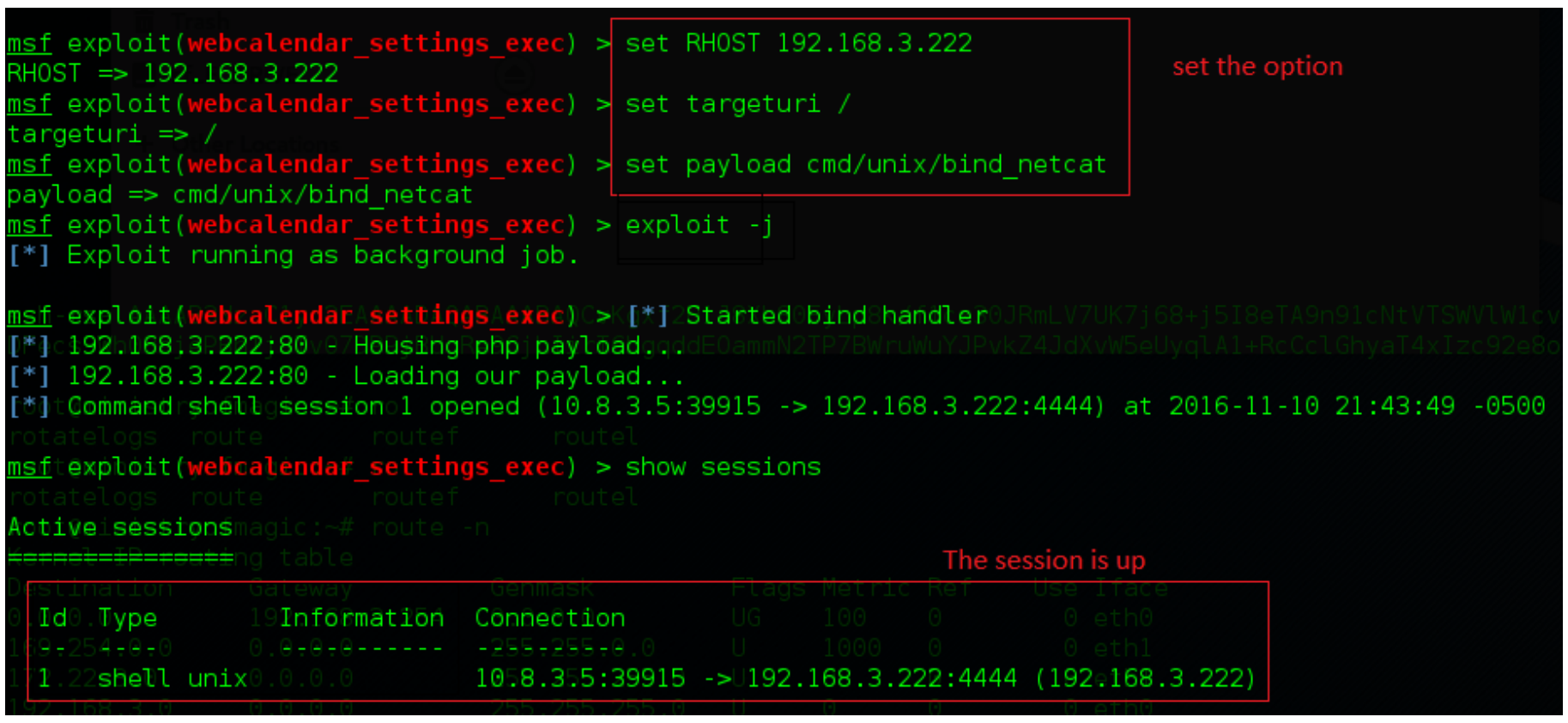

Figure 18. Run the exploit and get the session.

To upgrade a Normal Command Shell to a Metasploit Meterpreter, the upgrade module was identified by searching for shell_to_Meterpreter which was set in session one, as shown in Figure 19.

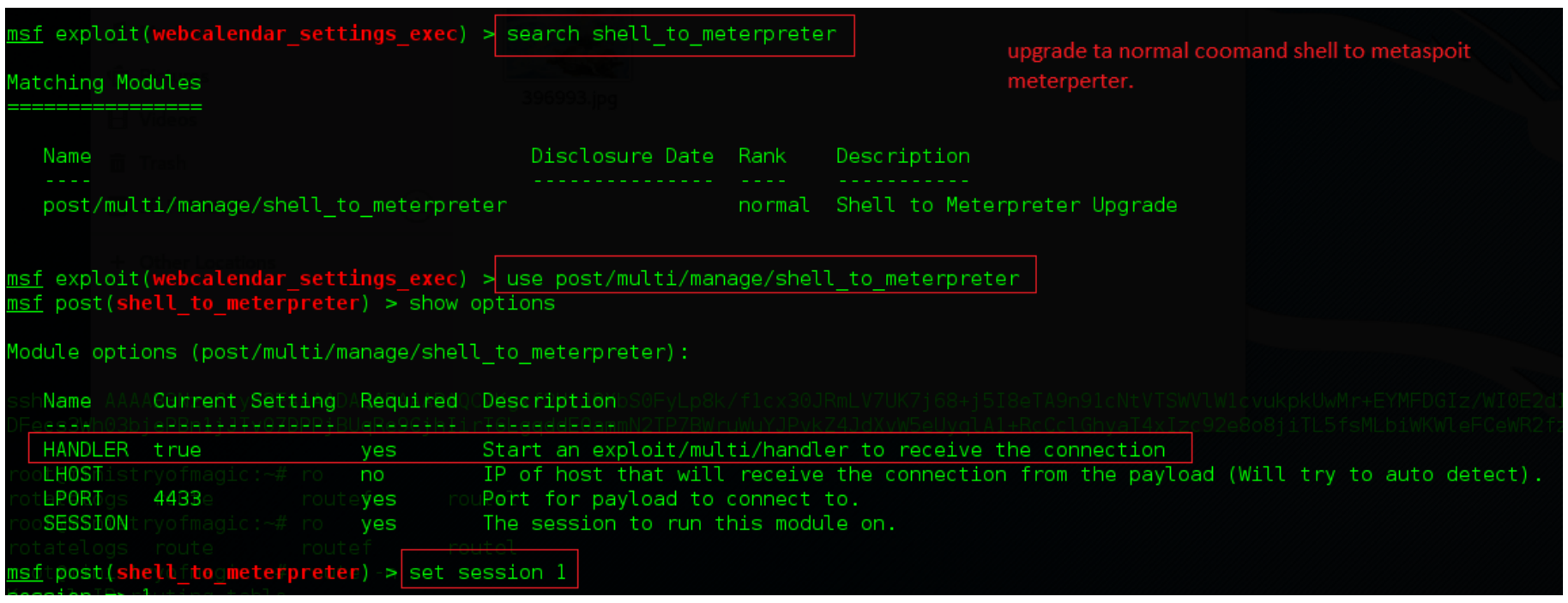

Figure 19. Search shell_to_meterpreter

After the module was executed, the session was changed to a new Meterpreter session, in which control was given to the team, as shown in Figure 20.

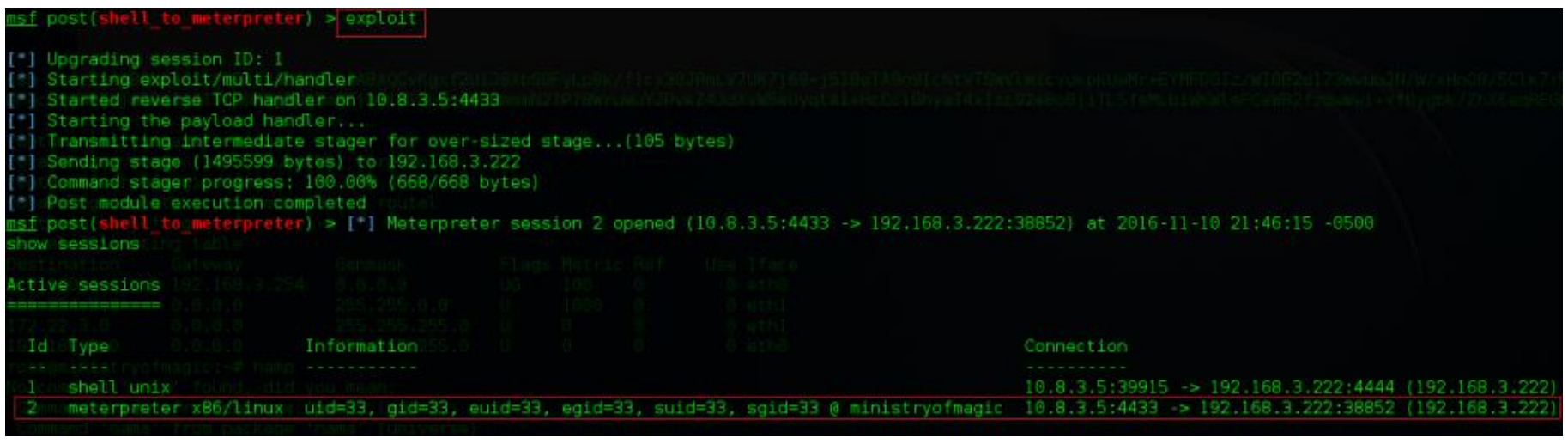

Figure 20. Run expolit shell_to_merepreter.

The meterpreter session was session 2. Therefore, to start interacting with it, the session's -i command was entered, followed by session number (2) to start the interaction, as shown in Figure 21.

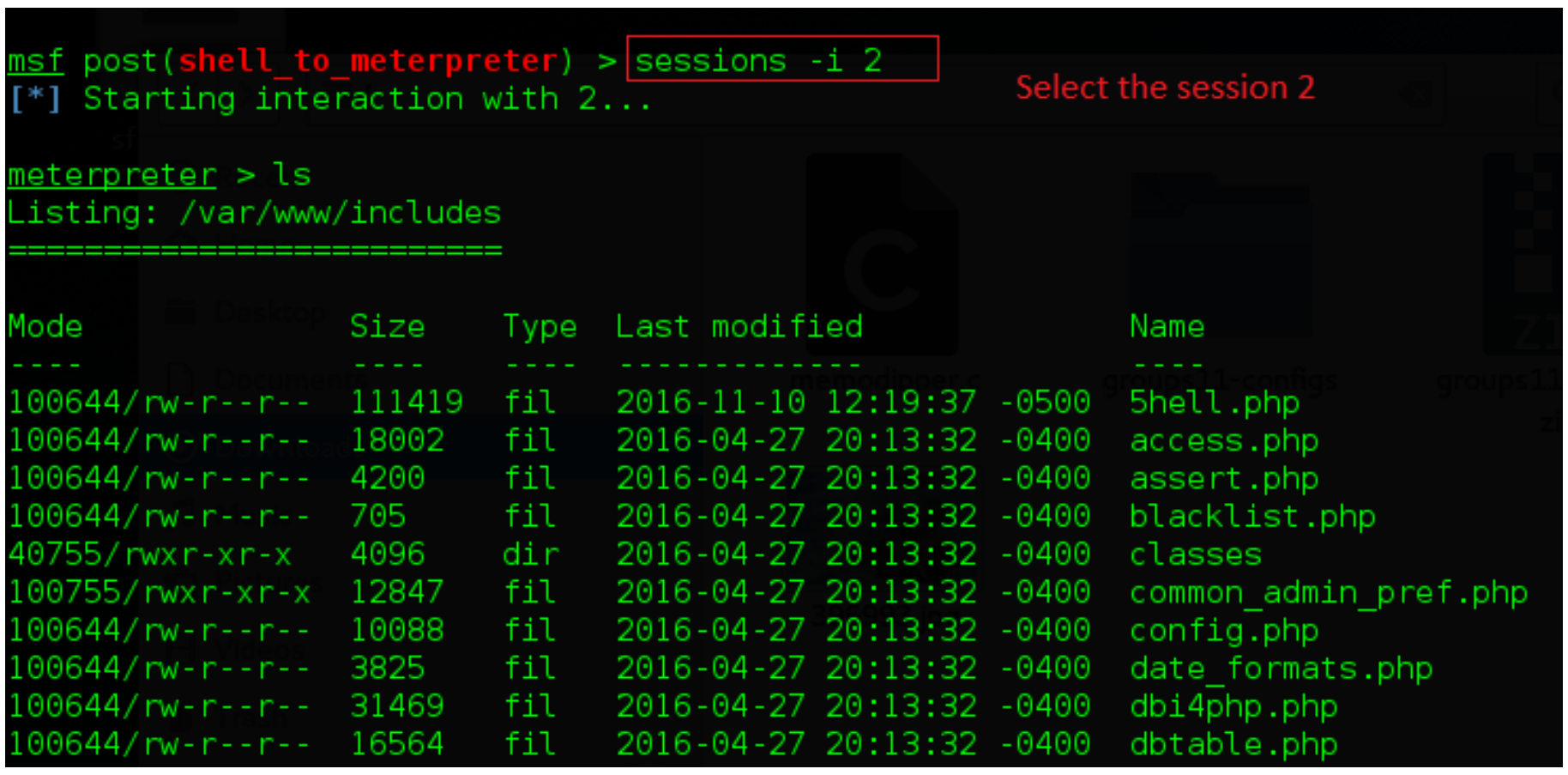

Figure 21. Starting with session two and using a command to list files.

To escalate to root privilege needs a code to run on the server-side. The code name "meodipper.c" was found. The next step was to update the file to the victim's server, as shown in Figure 22.

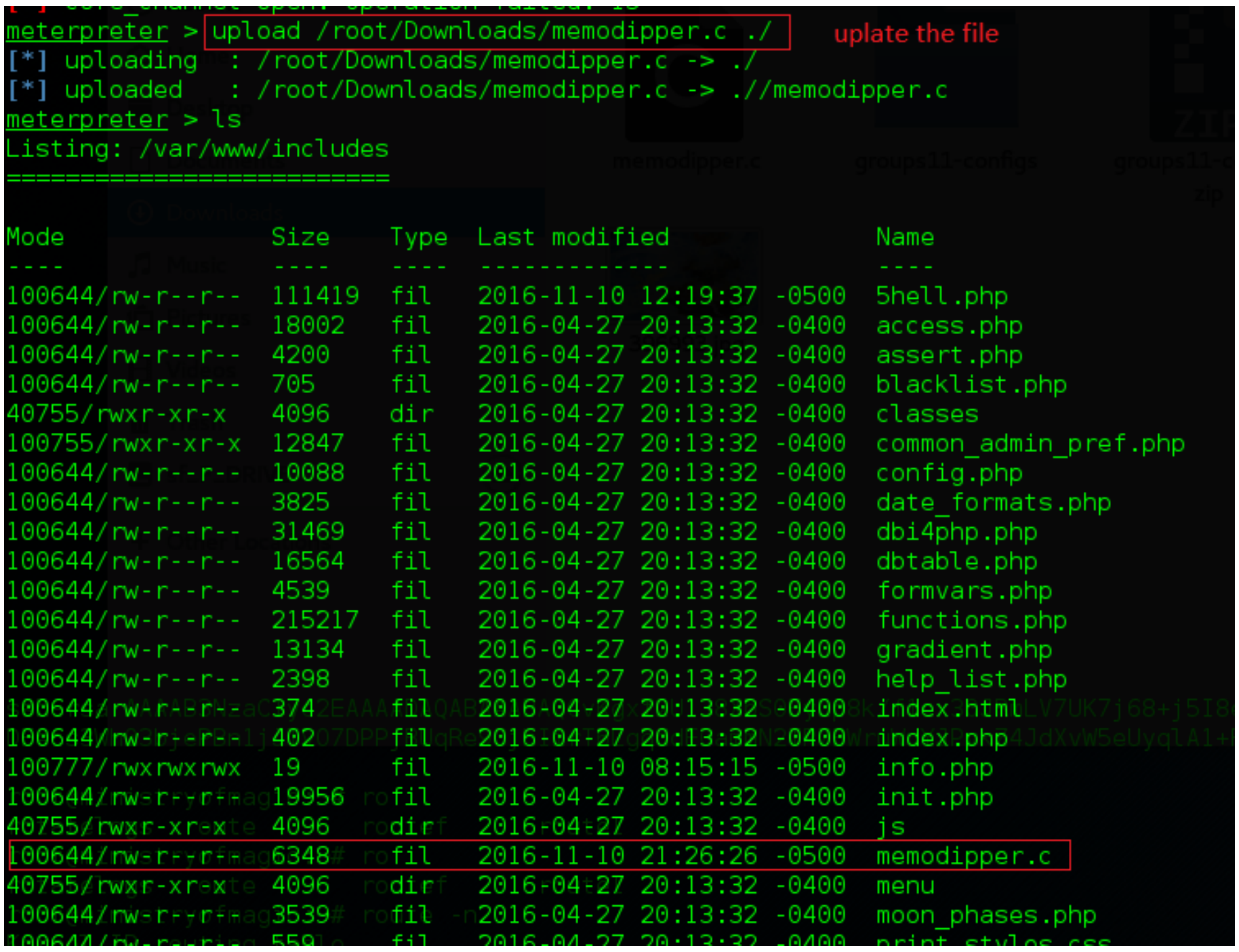

Figure 22. Upload The File Memodipper.c to server.

Since the code was written in the C language, the GCC Compiler was used. It is a very powerful and popular C compiler for various Linux [2]; commands and is helpful for the compiler of the code, as shown in Figure 23. In addition, it shows the following procedure was running a file to gain root privilege. That step reset the root password, as shown in Figure 24.

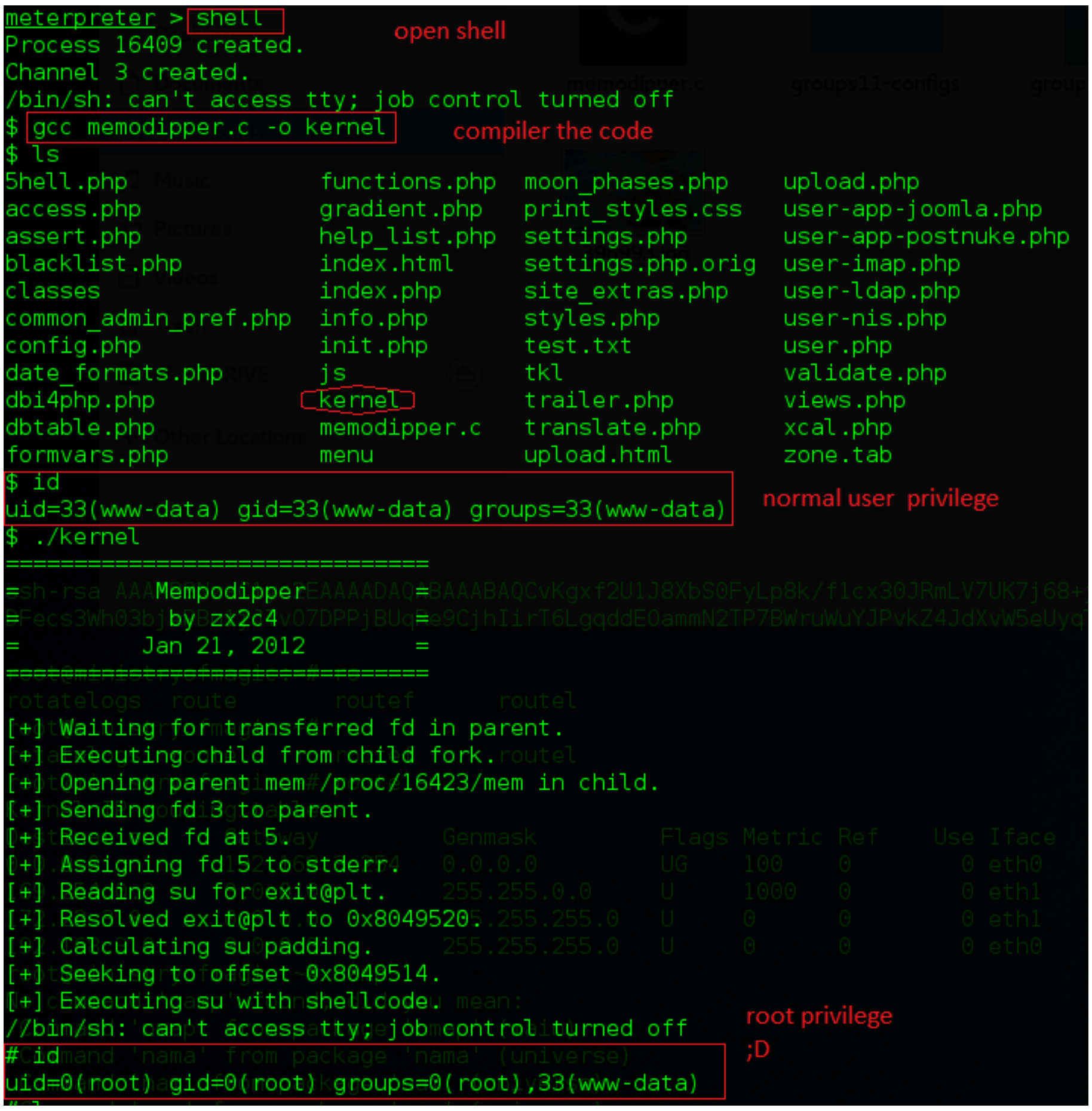

Figure 23. Escalate to root privilege.

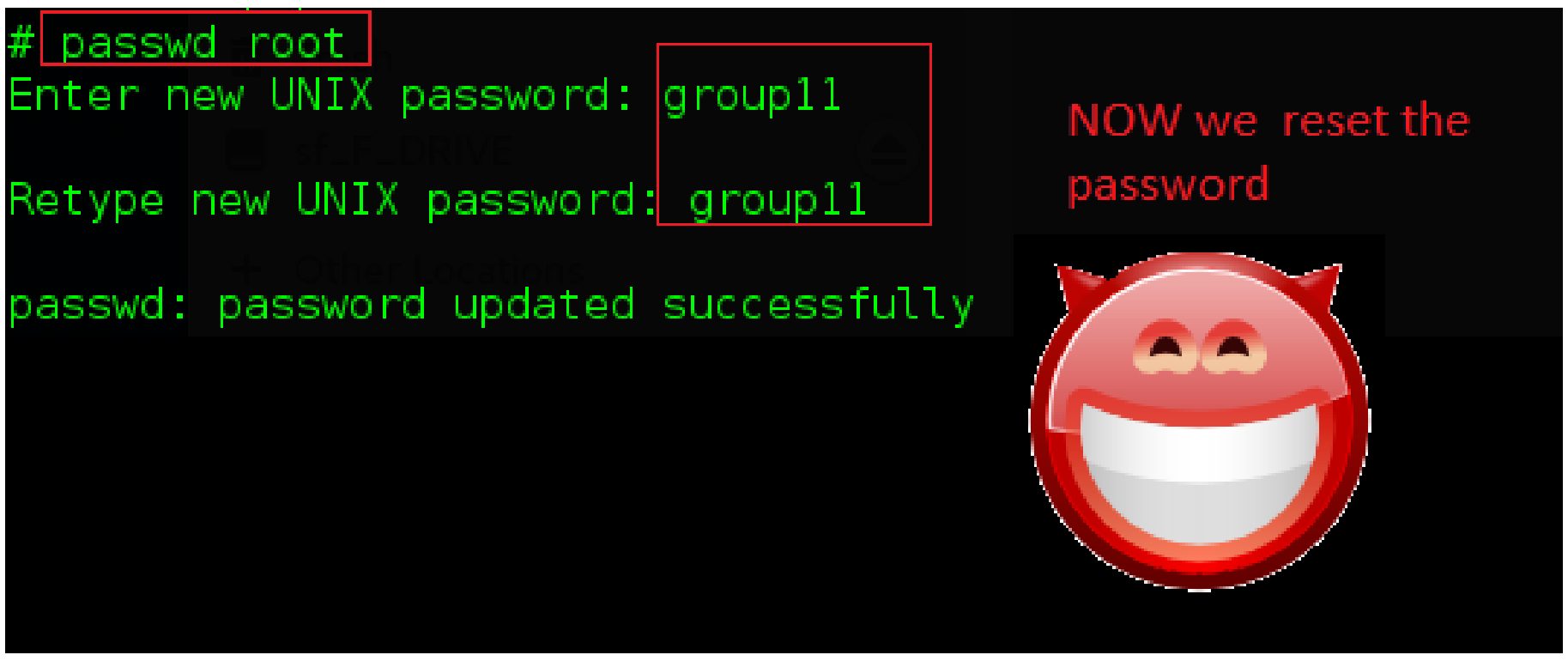

Figure 24. Reset the root password.

## 4) Flag Findings:

Once the system was successfully compromised, sophisticated Flags were searched for. Fortunately, two flags were found, the first being a flag.text under /home/snape/Document, as seen in Figure 25.

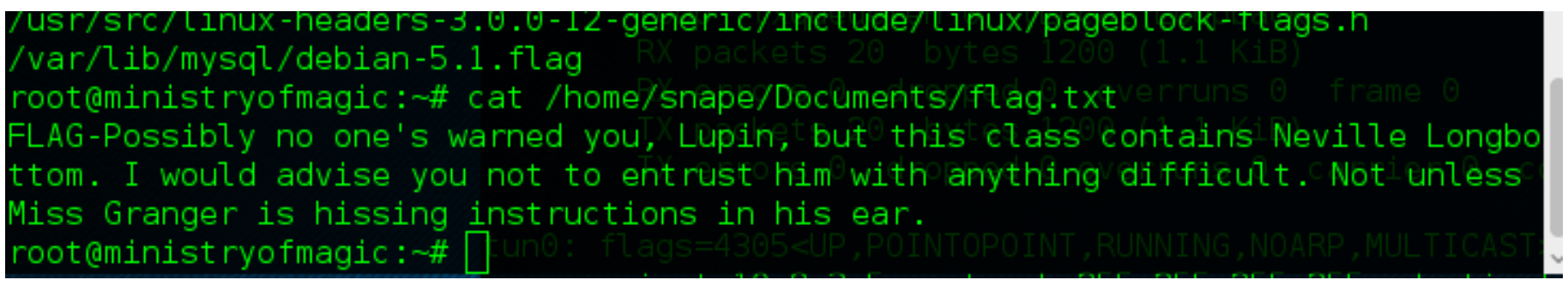

Figure 25. Flag 08.

The second Flag file was a Wireshark file named “flag.pcap” under /home/snape/Download. At many networks, packets were searched. For example, at HTTP packets were found with an interesting request from 192.168.247.1 host for Harry Potter. Thus, the Flags at frame sequence 4000 were detected, from source 192.168.247.98 to destination host 192.168.247.1, as shown in Figure 26. Fortunately, the second interesting flag was captured when the bytes of the packets were exported, as shown in Figure 27.

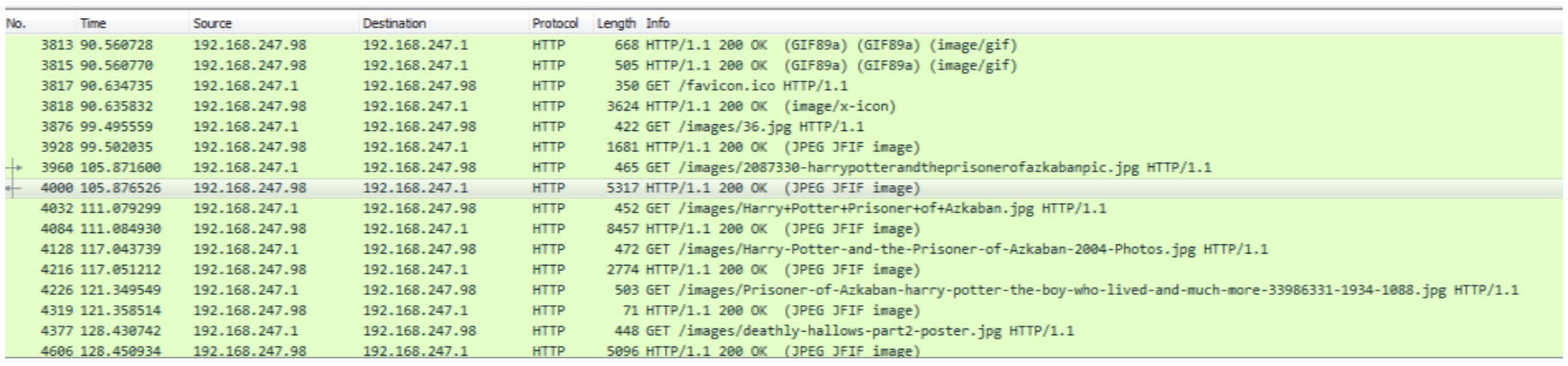

| No. | Time | Source | Destination | Protocol | Length | Info |
|---|---|---|---|---|---|---|
| 3813 | 90.560728 | 192.168.247.98 | 192.168.247.1 | HTTP | 668 | HTTP/1.1 200 OK (GIF89a) (GIF89a) (image/gif) |
| 3815 | 90.560770 | 192.168.247.98 | 192.168.247.1 | HTTP | 505 | HTTP/1.1 200 OK (GIF89a) (GIF89a) (image/gif) |
| 3817 | 90.634735 | 192.168.247.1 | 192.168.247.98 | HTTP | 350 | GET /favicon.ico HTTP/1.1 |
| 3818 | 90.635832 | 192.168.247.98 | 192.168.247.1 | HTTP | 3624 | HTTP/1.1 200 OK (image/x-icon) |
| 3876 | 99.495559 | 192.168.247.1 | 192.168.247.98 | HTTP | 422 | GET /images/36.jpg HTTP/1.1 |
| 3928 | 99.502035 | 192.168.247.98 | 192.168.247.1 | HTTP | 1681 | HTTP/1.1 200 OK (JPEG JFIF image) |
| 3960 | 105.871600 | 192.168.247.1 | 192.168.247.98 | HTTP | 465 | GET /images/2087330-harrypotterandtheprisonerofazkabanpic.jpg HTTP/1.1 |
| 4000 | 105.876526 | 192.168.247.98 | 192.168.247.1 | HTTP | 5317 | HTTP/1.1 200 OK (JPEG JFIF image) |
| 4032 | 111.079299 | 192.168.247.1 | 192.168.247.98 | HTTP | 452 | GET /images/Harry+Potter+Prisoner+of+Azkaban.jpg HTTP/1.1 |
| 4084 | 111.084930 | 192.168.247.98 | 192.168.247.1 | HTTP | 8457 | HTTP/1.1 200 OK (JPEG JFIF image) |
| 4128 | 117.043739 | 192.168.247.1 | 192.168.247.98 | HTTP | 472 | GET /images/Harry-Potter-and-the-Prisoner-of-Azkaban-2004-Photos.jpg HTTP/1.1 |
| 4216 | 117.051212 | 192.168.247.98 | 192.168.247.1 | HTTP | 2774 | HTTP/1.1 200 OK (JPEG JFIF image) |
| 4226 | 121.349549 | 192.168.247.1 | 192.168.247.98 | HTTP | 503 | GET /images/Prisoner-of-Azkaban-harry-potter-the-boy-who-lived-and-much-more-33986331-1934-1088.jpg HTTP/1.1 |
| 4319 | 121.358514 | 192.168.247.98 | 192.168.247.1 | HTTP | 71 | HTTP/1.1 200 OK (JPEG JFIF image) |
| 4377 | 128.430742 | 192.168.247.1 | 192.168.247.98 | HTTP | 448 | GET /images/deathly-hallows-part2-poster.jpg HTTP/1.1 |
| 4606 | 128.450934 | 192.168.247.98 | 192.168.247.1 | HTTP | 5096 | HTTP/1.1 200 OK (JPEG JFIF image) |

Figure 26. Wireshark harry potter reply image “Flag”.

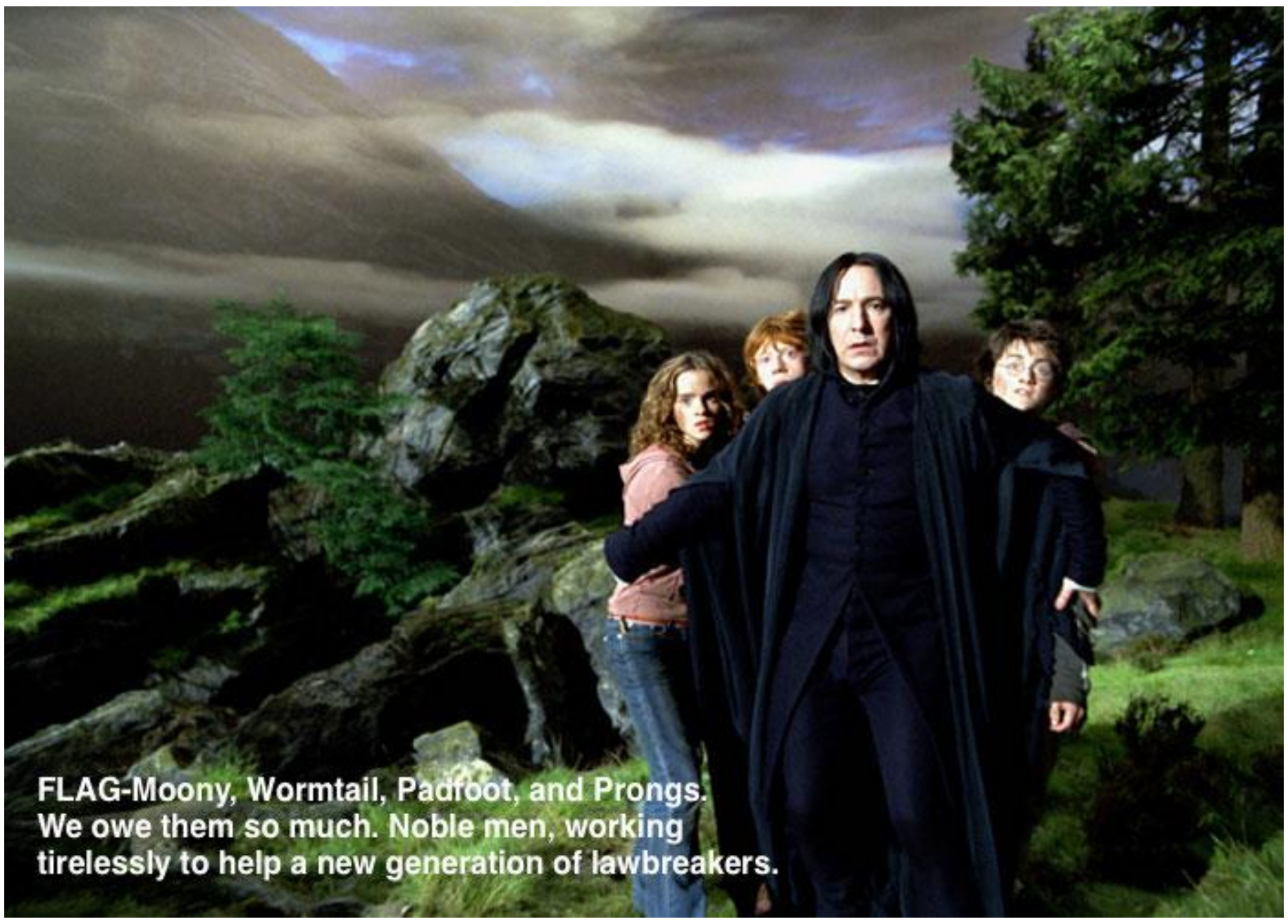

Figure 27. Flag text message.

# IV. Mitigation and Recommendations

In this paper, we described how different systems were compromised regarding different service vulnerabilities. To mitigate such risks, the defences against cyber-attack must be highly prioritised for any network systems. This paper highlights various mitigation strategies to reduce system vulnerability risks found in this network.

## 1) FTP the Achilles Heel:

As a company uses FTP to transfer information over the network from one machine to another, there is always a risk of data breach, causing millions of customers’ data to be exposed to hackers. The worst part is that the data breach cannot be easily traced as the logs files can be manipulated by unauthorised use.

There are some questions that every administrator should ask:

- How many machines are using FTP services?
- Which FTP machine is utilised most?
- Which is the most used FTP user account?
- Which files are most utilised?
- Are the FTP services up to date?

Identifying FTP user’s recommendations:

The first step is to identify anonymous users and close those accounts. The next the step is to detect rogue traffic coming in and out of the machines. This can be difficult if the user accounts do not have a centralised login system. This is usually not the case because mostly FTP accounts are set up per machine basis as a temporary solution, making it difficult for an administrator to monitor traffic coming in and out of the machines. The administrators should use a red flag if an anonymous login to FTP services seems to result in a spike in services. If unknown accounts surface during the monitoring phase, this should raise a high alert for administrators and inquiries should be made.

### 2) SSH:

It is one of the ubiquitous methods for working with networked machines, and it is also a fact that many hackers try to exploit this method to gain access to a machine [5]. This can be addressed as follows:

- Limiting the number of attempts.
- Strict rules should be implemented for access.
- Keeping track of IPs that try to brute force and deny access.
- Rotate the passwords of the SSH accounts to minimise unauthorised access.
- Disable access with SSH keys to the machines which contain sensitive information.
- In no situation should an SSH account with root access be allowed to be used with the SSH protocol.
- Keeping track of when and who accesses the specific machine with SSH.

Additionally, the network administrator should be fully aware of FTP service vulnerabilities [5], and different security policies should be considered, such as:

- Restricting FTP server access based on network IP addresses.
- Limiting the sending password attempts to prevent brute force passwords.
- Applying password policy.
- Using an authentications mechanism which is not subject to eavesdropping while sending passwords.
- Applying appropriate updates.
- Minimising the number of users with admin privileges.
- Using a MFT (Managed Files Transferred) server for transferring data to provide secure internal and external file transfer rather than standard FTP.
- Remote privileged access must also be monitored to detect any attack attempts. In addition, security measures such as multi-factor authentication, login location and time-of-day restrictions should be seriously considered.
- Keep updating any FTP & HTTP utilities and avoid any known vulnerable versions.
- Disabling anonymous accounts for FTP default access or preventing anonymous users from uploading files services to restrict Write Access for those users.
- Use an SFTP (SSH File Transfer Protocol) instead of standard FTP to provide secure file transfer.
- Implementation of change control for all systems: misconfiguration, insecure deployment and keeping systems outdated were explored among various systems. The vulnerabilities can be mitigated by enforcing change control processes on all systems.
- Conducting a regular vulnerability assessment as one effective company risk management strategy to evaluate if the installed security controls are correctly installed.
- Implementing system monitoring and detection systems to detect situations before they become incidents.

## V. Conclusion

To conclude, in this paper penetration testing objectives have been met with various system vulnerabilities, resulting in a complete compromise of some hosts. Consequently, eight critical data "Flags" were collected. Furthermore, the systems had many issues, which may typically be expected minor issues. However, these resulted in compromising the victims and gaining access to critical data. Therefore, it is highly recommended that the suggested mitigation strategies be applied to protect against any future attacks.